\documentclass[]{article}

\usepackage[utf8]{inputenc}

\usepackage{cite}

\usepackage{nameref,hyperref}

\usepackage[right]{lineno}

\usepackage{microtype}
\DisableLigatures[f]{encoding = *, family = * }

\raggedright
\usepackage{changepage}

\usepackage[aboveskip=1pt,labelfont=bf,labelsep=period,singlelinecheck=off]{caption}

\makeatletter
\renewcommand{\@biblabel}[1]{\quad#1.}
\makeatother

\usepackage{lastpage,fancyhdr,graphicx}
\usepackage{epstopdf}
\fancyheadoffset[L]{2.25in}
\fancyfootoffset[L]{2.25in}

\usepackage{color}

\definecolor{Gray}{gray}{.25}

\usepackage{graphicx}

\usepackage{sidecap}

\usepackage{wrapfig}
\usepackage[pscoord]{eso-pic}
\usepackage[fulladjust]{marginnote}
\reversemarginpar

\begin{document}
\vspace*{0.35in}

\begin{flushleft}






{\Large
 \textbf\newline{Long-term changes in functional connectivity predict responses to intracranial stimulation of the human brain}
 } %

\bigskip



\bigskip

Christoforos A Papasavvas\textsuperscript{1},
Peter Neal Taylor\textsuperscript{1,2,3},
Yujiang Wang\textsuperscript{1,2,3,*}
\\
\bigskip

\begin{enumerate}
\item{Interdisciplinary Computing and Complex BioSystems Group, School of Computing, Newcastle University, Newcastle upon Tyne, United Kingdom}
\item{Faculty of Medical Sciences, Newcastle University, Newcastle upon Tyne, United Kingdom}
\item{UCL Queen Square Institute of Neurology, Queen Square, London, United Kingdom}
\end{enumerate}

\bigskip
*corresponding author: Yujiang Wang, yujiang.wang@ncl.ac.uk

\bigskip
Conflict of interest statement: The authors declare no competing interests.

\subsubsection*{Acknowledgments}
We thank the CNNP team (www.cnnp-lab.com) for their comments on the manuscript. CAP, PNT, and YW gratefully acknowledge funding from Wellcome Trust (208940/Z/17/Z and 210109/Z/18/Z).

Data were provided in part by the Human Connectome Project, WU-Minn Consortium (Principal Investigators: David Van Essen and Kamil Ugurbil; 1U54MH091657) funded by the 16 NIH Institutes and Centres that support the NIH Blueprint for Neuroscience Research; and by the McDonnell Centre for Systems Neuroscience at Washington University.

\end{flushleft}

\clearpage

\section*{Abstract}
Targeted electrical stimulation of the brain perturbs neural networks and modulates their rhythmic activity both at the site of stimulation and at remote brain regions. Understanding, or even predicting, this neuromodulatory effect is crucial for any therapeutic use of brain stimulation. To this end, we analyzed the stimulation responses in 131 stimulation sessions across 66 patients with focal epilepsy recorded through intracranial EEG (iEEG). We considered functional and structural connectivity features as predictors of the response at every iEEG contact. Taking advantage of multiple recordings over days, we also investigated how slow changes in interictal functional connectivity (FC) ahead of the stimulation relate to stimulation responses. The results reveal that, indeed, this long-term variability of FC exhibits strong association with the stimulation-induced increases in delta and theta band power. Furthermore, we show through cross-validation that long-term variability of FC improves prediction of responses above the performance of spatial predictors alone. These findings can enhance the patient-specific design of effective neuromodulatory protocols for therapeutic interventions.



\section*{Introduction}

Direct electrical brain stimulation constitutes an increasingly useful therapeutic intervention for neurological, neuropsychiatric, and neurocognitive disorders \cite{Coubes2000, Fisher2010, Koller1997, Benabid1991, Holtzheimer2017, Suthana2012, Mayberg2005, Kisely2018}. It has been successfully used for tremor treatment in Parkinson's disease \cite{Koller1997, Benabid1991} and promising results have been reported for the treatment of severe depression \cite{Mayberg2005, Kisely2018}. While the mechanisms behind the beneficial effect of stimulation are only partially understood \cite{Herrington2016}, the stimulation protocol is typically designed through a process of trial and error \cite{Volkmann2002}. However, in order to maximize its therapeutic potential more research is needed to reliably predict the stimulation effect on an individual level, so that we can design optimal stimulation protocols for each patient.

Multiple recent and ongoing investigations have been focused on revealing the various predictive or influencing factors of the stimulation effect. Stimulation parameters, such as amplitude and frequency, can cause both increase and decrease of gamma activity in responding brain areas, depending on the specific parameters used \cite{Mohan2020}. The proximity of stimulation to anatomical elements, such as white-matter tracts, has also been reported as a factor which correlates with the stimulation-induced increase in theta and gamma band power  \cite{Solomon2018, Mohan2020}. Interestingly, network interactions may also be relevant; the beta band activity in a \textit{modulator} area of a network has been shown to predict the stimulation-driven communication between two other areas in the same network \cite{Qiao2020}. In addition, a great interest has been developed on the association between functional connectivity and stimulation responses. Multiple sources of evidence suggest that brain areas respond stronger to stimulation if they are strongly functionally connected to the stimulation site \cite{Solomon2018, Huang2019}. In particular, the functional network of the medial temporal lobe has been shown to correlate with the increases in theta band power as a response to stimulation \cite{Solomon2018}. Predictability of theta band neuromodulation through stimulation is particularly important due to the functional relevance of theta oscillations in memory and other cognitive processes (for reviews see \cite{Buzsaki2013, Herweg2020}).

Furthermore, the research community has started exploring the slow changes of functional brain networks over days and weeks, and how they relate to either ongoing physiological rhythms or fluctuations of the disease \cite{Schroeder2020, Mitsis2020}. Such studies are increasingly supported for datasets that provide long-term recordings spanning multiple days. Since slow functional reorganization has been associated with increased sensitivity to stimulation \cite{Verley2018}, we hypothesized that slow changes in functional connectivity could potentially explain the stimulation neuromodulatory effects, something which has been unexplored until now.

Here we investigated whether various functional and structural connectivity measures between the stimulation site and other brain areas can predict the level of responses in the corresponding brain areas. Taking advantage of a long-term dataset including 66 subjects, we introduce a measure that quantifies the variability of functional connectivity across days. We corroborate the previous findings on the association between functional connectivity and theta modulation \cite{Solomon2018}, and show additionally that indeed the long-term variability of functional connectivity substantially improve predictive performance for the stimulation responses in delta and theta band power.

\section*{Materials and Methods}

\subsection*{Experimental Design and Statistical Analysis}
The present study is based on previously acquired data. The most important elements of the experimental design are noted here, but for further details, see the original studies \cite{Solomon2018, Ezzyat2018}. Heterogeneous statistical analyses and models were used throughout this study as detailed in the following subsections in Materials and Methods. 

\paragraph*{Electrophysiological recordings}

We used openly available data from Phase I and Phase II (up to Year 3) of the Restoring Active Memory (RAM) project (University of Pennsylvania; \url{http://memory.psych.upenn.edu/RAM}). Patients with drug-resistant epilepsy were recruited during their clinical treatment with intracranial electroencephalographic monitoring. As stated in the project's website "Informed consent has been obtained from each subject to share their data, and personally identifiable information has been removed to protect subject confidentiality". The original data acquisition was approved by the University of Pennsylvania, while the University Ethics Committee at Newcastle University approved our data analysis (Ref: 12721/2018). 

We considered 66 out of the total 79 subjects for which we calculated the stimulation responses in our previous study \cite{Papasavvas2020}. We excluded 13 subjects from the present study by applying two criteria. Since we were interested in the long-term, cross-session FC changes, we excluded those who had no stimulation-free sessions before their first stimulation session. We also excluded those who had less than 30 recorded channels (not counting anode/cathode). This exclusion was applied for statistical reasons (see below for the definition of highly responding sessions based on the distribution of responses). The remaining 66 subjects had 131 stimulation sessions in total and each subject had at least one stimulation-free session preceding their first stimulation session.

\paragraph*{Stimulation paradigm}

Charge-balanced biphasic rectangular pulses with 300 $\mu$s width were used for stimulation through neighboring bipolar anode/cathode electrodes \cite{Solomon2018}. Despite the bipolar nature of stimulation, anodes and cathodes were annotated separately in the metadata and we considered that annotation in our analysis. The stimulation was delivered with frequency 10, 25, 50, 100, or 200 Hz. Its amplitude varied from 0.25 to 3.5 mA and the duration of the pulse train was 500 ms or 4.6 s, depending on the experiment. The pulse train duration did not vary between sessions within a subject.

\subsection*{Measure of stimulation response}

The neuromodulatory effect of stimulation was measured exactly as in our previous work \cite{Papasavvas2020}. Here follows a summary of the most important steps. We extracted 1-second iEEG segments just before and just after each stimulation trial, which represent the pre and post stimulation activity, respectively. After applying common average re-referencing, we quantified the band power of each segment in five frequency bands: delta (2–4 Hz), theta (4–8 Hz), alpha (8–12 Hz), beta (12–25 Hz), and gamma (25–55 Hz). The band powers were log-transformed before quantifying the pre-to-post neuromodulatory effect of stimulation (i.e. stimulation responses). This quantification was achieved by using the z-statistic of a paired Wilcoxon sign rank test applied to the paired band power values, pre vs post, across the multiple trials during the session.

\subsection*{Functional connectivity measures}

A 60-second segment was extracted from the beginning of each stimulation and stimulation-free session. These initial segments were never contaminated with stimulation trials. Each segment was split into 30 equal-width non-overlapping blocks as shown in Fig.~\ref{fig1}a. The broadband functional connectivity (FC) for all channel pairs was calculated using Pearson correlation for each one of these blocks and then an FC matrix was constructed based on the average FC across the blocks. Another FC matrix was constructed based on the standard deviation of FC across the blocks (see Fig.~\ref{fig1}). 

Separate functional connectivity measures of anode and cathode, $FC_A$ and $FC_C$, were used. These measures were extracted from the corresponding rows in the average FC matrix, as shown in Fig.~\ref{fig1}b. A derivative measure, $FC_M$, was also computed by mixing the FC information from anode and cathode. The $FC_A$ and $FC_C$ values were first z-scored separately and then the average absolute value was taken for each channel; thus producing $FC_M$ as shown Fig.~\ref{fig1}b.

In an analogous way, three additional measures were produced but derived from the standard deviation FC matrix instead. The short-term variability of FC for anode and cathode, $FC_{sA}$ and $FC_{sC}$, was taken directly from the corresponding rows in the standard deviation FC matrix, as shown in Fig.~\ref{fig1}c. An anode/cathode mixed measure of short-term variability, $FC_{sM}$, was also computed, again by z-scoring separately $FC_{sA}$ and $FC_{sC}$ and then taking the average absolute value for each channel.

The long-term variability of FC, $FC_L$, is the only measure that extracts FC information from multiple sessions and more specifically from their average FC matrices. Z-scored $FC_A$ and $FC_C$ are computed not only from the beginning of the to-be-predicted stimulation session but also from all its preceding sessions. The session timestamps are provided thus the ordering of the sessions for each subject is straightforward. For every stimulation session there is at least one preceding session and at least the first of those preceding sessions needs to be a stimulation-free session (see above for subjects considered in the study). All the available z-scored $FC_A$ and $FC_C$ are aligned and the measure $FC_L$ is computed as the channel-wise standard deviation as shown in Fig.~\ref{fig1}d. $FC_L$ expressed the variability of FC over multiple sessions, considering simultaneously anode and cathode FC.

\subsection*{Structural connectivity measures}
Electrode contacts were localized to individual brain regions using the publicly available MNI space coordinates. We first used surface based registration of the Lausanne 250 parcellation \cite{Hagmann2008, Daducci2012} to the MNI brain with the \texttt{mri\textbackslash\_surf2surf} command in FreeSurfer \cite{Fischl2012}. Volumetric ROIs for the Lausanne 250 parcellation were then generated using the \texttt{mri\textbackslash\_aparc2aseg} command. Next, the volumetric ROIs were imported into MATLAB, along with the coordinates of the electrode contacts, and in house code calculated the minimum Euclidean distance of each contact to each ROI voxel. The closest ROI was then assigned to that electrode contact.

To construct a representative structural connectivity matrix we used  a group average template, constructed from a total of 1065 subjects from the Human Connectome Project \cite{Van2013}. A multishell diffusion scheme was used, with b-values set to 990, 1985 and 2980 s/mm\textsuperscript{2} and 90 diffusion sampling directions per shell. Data were acquired with a 1.25 mm isotropic voxel size. The b-table was checked by an automatic quality control routine to ensure its accuracy \cite{Schilling2019}. The diffusion data were reconstructed in the MNI space using q-space diffeomorphic reconstruction \cite{Yeh2011}.  A diffusion sampling length ratio of 1.7 was used and the output resolution was 1 mm isotropic. Deterministic fiber tracking was then performed using whole brain seeding until 10\textsuperscript{7} tracts were found within specific criteria \cite{Yeh2013}. Criteria included the anisotropy threshold, which was randomly selected, and the step size which was randomly selected from 0.5 to 1.5 voxels. The angular threshold was randomly selected from 15 to 90 degrees and tracks with length shorter than 10 or longer than 300 mm were discarded. Topology-informed pruning \cite{Yeh2019} was applied to the tractography with 1 iteration to remove false connections.  Volumetric ROIs from the MNI space Lausanne 250 parcellation were used as terminating regions, and two connectivity matrices were calculated as the number of connecting streamlines (streamline count, $SC_{\textit{slc}}$, log-transformed using $\log_{10}(n+1)$) and mean generalized fractional anisotropy of tracts connecting regions ($SC_{\textit{gfa}}$).

\subsection*{Inverse Euclidean distance}

We considered the inverse Euclidean distance between stimulation site and response site, $D = 1/(\textrm{Euclidean distance in mm})$, as another feature for our modeling. The stimulation site in this case was considered to be the midpoint between anode and cathode.

\subsection*{Definition of subsets of highly responding sessions}

A subset of highly responding sessions was formed for each frequency band. A highly responding session for a specific band is considered to be a session with a 95\textsuperscript{th} percentile of at least 2.4 in the distribution of response values, across its channels, for the specific band. The subsets of highly responding sessions are thus different for each band. The rational behind using these subsets is that any successful attempt for response predictions will be made on sessions with at least a few channels that distinguish themselves from the baseline fluctuations of band power. The threshold of 2.4 was chosen, considering that 95\% of the baseline fluctuations range from -1.98 to 2.16 across all bands (for distributions, see \cite{Papasavvas2020}).

\subsection*{Linear mixed-effects models}

The linear mixed-effects models used had in general the following form: $R \sim 1 + X_1 + ... + X_q + (1|S)$. The dependent variable R represents the stimulation responses, that is, modulation of band power in a particular frequency band. A linear combination of one or more independent variables $X$ were used to model the response $R$, while the features defined above served as the independent variables. The channels were grouped per session, indicated here as categorical variable $S$, so that a random intercept was allowed for each session, as indicated by the term $(1|S)$. A random intercept could account for any systematic variation of the average response from session to session. In contrast, we decided to keep the slope fixed so that it could capture common associations across the sessions. The prediction relies heavily on such common associations (see below).

Before fitting the models, we always applied standardization of each feature within a session. We used MATLAB (R2019a) function \texttt{fitlme} with Maximum Likelihood Estimation as the fitting method. After fitting, the resulting parameter estimates for each independent variable was divided by its standard error (as in \cite{Mohan2020}). We consider this as an effect measure, termed here as associative power (AP). Note that, throughout the paper, we used effect measures rather than p-values. We intentionally avoided significance test for reporting presence/absence of effects. Some p-values are reported only for reference; we did not use them for downstream analysis or as exclusion/inclusion criteria.

\subsection*{Bootstrapping}

We used bootstrapping to produce all 95\% confidence intervals used. We generated 2000 resamples of sessions/channels by randomly sampling with replacement. The size of the resamples always matched the size of the corresponding sets of sessions on which bootstrapping was applied. Those were either the whole set of 131 sessions or band-specific subsets of highly-responding sessions. Model fitting was then applied to each resample separately.

\subsection*{Step-wise analysis and statistical comparison of models}

We applied both forward and backward step-wise analysis on multivariate mixed-effects models. In forward analysis, the most enriching feature was added to the model in each step, based on statistical comparison of models (see below). In backward analysis, the least informative feature already in the model was removed in each step, using the same statistical comparison.

For the comparison of models we used the Likelihood Ratio test as implemented in MATLAB (R2019a) (\texttt{compare} function in the class \texttt{LinearMixedModel}). The comparison was always carried out between two models where one model was nested in the other, that is, the set of features in one model was a subset of features used in the other model. The use of this statistical test reveals whether any additional features in the extended model can substantially enrich the nested model.

\subsection*{Prediction process and its assessment}
We systematically trained the models for prediction using three different cross-validation schemes and we assessed the prediction as follows. We used MATLAB (R2019a) function \texttt{predict}. First, for the prediction of the response at a particular channel during a particular session, we trained the models on the rest of the channels and sessions. The training set included not only the remaining channels of the session, but also the channels in all the other sessions. We assessed the goodness of prediction by computing the squared error between the true response and the predicted one for each individual channel. Second, we predicted subsets of channels in each session; more specifically, half of the channels. Since there are many possible combinations of channels, we applied prediction on 500 randomly chosen subsets of channels in each session. In every case, we trained the models on the remaining channels across sessions. We assessed the goodness of prediction by computing the Pearson correlation, $r$, between the predicted responses at the specific channels and their true values. Third, we applied prediction on whole sessions, that is, all the channels in a session. The training relied on the remaining sessions. Once again, the assessment of the prediction was based on the Pearson correlation between the true and predicted responses. 

By using Pearson correlation, we captured the agreement in channel ordering between prediction and empirical values, but also the relative distances between them. We also took advantage of the insensitivity of Pearson correlation to any scaling differences between the two distributions.

\subsection*{Measures of binary prediction}

After predicting all the channels in a particular session, we also assessed whether we can correctly predict the Y\% most responsive channels in that session. Each channel was labeled as highly responding channel if it was predicted to be in the top Y\% of channels in the session in terms of responsiveness. Otherwise, it was labeled as weakly responding channel. We assessed the quality of this classification by using two measures: area under the receiver operating characteristic curve (AUC) and sensitivity. Sensitivity, or true positive rate, is the ratio of true positives, correctly labeled as highly-responding, over the x\% of channels in the session. We computed the AUC, which expresses the ability to discriminate between weakly- and highly- responding channels, by using \texttt{perfcurve} in MATLAB (R2019a).  

For the percentage of top channels, we used values from 5\% to 30\%. We limit the analysis in this range due to the across-session distribution of channel percentages of highly-responding channels; those with stimulation responses of at least 2.4 (same threshold used for the definition of highly responding sessions, see above). For instance, the corresponding distribution for delta has a median of 11.0\% and interquartile range [7.7\% 23.8\%]. The corresponding distribution for theta has a median of 9.1\% and interquartile range [6.3\% 14.0\%].

\subsection*{Improvement of prediction}

To detect any prediction improvement after adding a feature to a model, we used a paired Wilcoxon sign rank test. The paired nature of the test is appropriate since the prediction was assessed using two models (i.e. paired assessments), with one being a nested model to the other. This statistical test for improvement was applied to all the different measures we used for the assessment of prediction: squared error, Pearson correlation, AUC, and sensitivity. We used primarily the z-statistic produced as the statistical effect representing improvement, but we also mention the corresponding p value for reference.

\subsection*{Code Accessibility}

The custom code producing all the results can be found online: [pending].

\section*{Results}

\begin{figure}[p]
\captionsetup{width=1.4\textwidth}
\hspace{-45mm}\vspace{0mm}\includegraphics{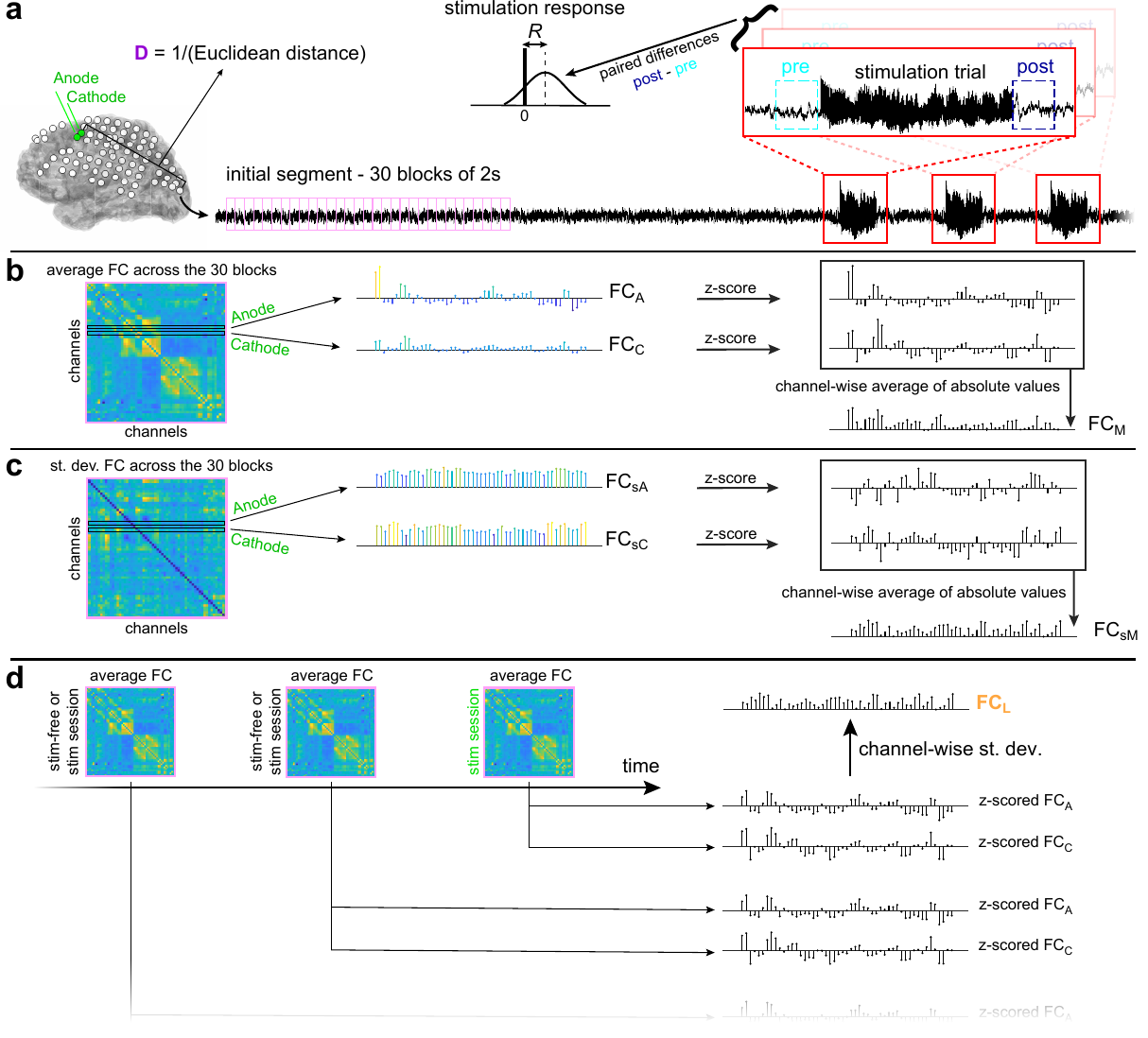}
\caption{
{\bf Functional connectivity features used for the modeling of stimulation responses.}
(a) In each stimulation session and for each response channel, the band-specific stimulation response was calculated by considering the paired differences of band power between pre and post across stimulation trials. Functional connectivity (FC) measures were based on 60s-long initial segments, split in 30 blocks without overlap. (b-c) FC between each pair of channels was calculated in each block and then the mean and standard deviation FC matrices were produced by applying mean and standard deviation across the blocks. FC measures of anode (A), cathode (C), and their mixture (M) were extracted from both average and standard deviation matrices. (d) Long-term variability of FC ($FC_L$) is the only feature that expresses the variability of FC across multiple sessions, achieved by applying channel-wise standard deviation across the sessions.
}
\label{fig1}
\end{figure}

Each subject in the study was implanted with intracranial EEG comprised of surface or depth electrodes, or a combination of both. Direct electrical stimulation was applied during some of the recording sessions. All electrode contacts recorded brain activity and during each stimulation session a pair of electrode contacts were used as the stimulation anode and cathode (Fig.~\ref{fig1}a). Multiple trials of stimulation were delivered during each stimulation session, and the stimulation responses were calculated by considering all available pairs of pre- and post-stimulation EEG segments (Fig.~\ref{fig1}a). We separated the responses into power modulations of the oscillatory activity in the five main frequency bands: delta, theta, alpha, beta, and gamma. The goal of this work is to test whether we can predict these stimulation responses using subject-specific data features before the stimulation.

The data features express electrophysiological, anatomical, and spatial relations between the stimulation and recording sites. As previous studies suggested, such relations have the potential to predict stimulation responses in specific frequency bands \cite{Solomon2018, Mohan2020}. Figures~\ref{fig1}b-d summarize the functional connectivity (FC) features used. The FC of the anode and cathode, $FC_A$ and $FC_C$, express the correlation between the recorded activity at the stimulation electrode contacts and every other recorded channel. Note that we used the initial period of the recording for the FC calculation, before any stimulation was performed. We further defined a feature that mixes the FC values of anode and cathode, $FC_M$, as a representative FC feature for the channel pair. We also defined features which express the temporal variability of FC during the initial period of the recording in each session. The defined features cover the FC variability at anode, $FC_{sA}$, at cathode, $FC_{sC}$, and a mixture of those, $FC_{sM}$. Finally, we defined a feature that expresses the long-term variability of FC, $FC_L$, based on the FC values of both anode and cathode over multiple past sessions. This last feature is the only one that depends not only on the current, to-be-predicted session, but also on preceding sessions.

Additional spatial and structural features were used, including the inverse Euclidean distance, $D$, as shown in Fig.~\ref{fig1}a, and two structural connectivity (SC) measures which carry information about the anatomical white matter connectivity between the stimulation area and the responding areas. We used the streamline count, $SC_{\textit{slc}}$, and generalized fractional anisotropy, $SC_{\textit{gfa}}$. Thus, we collated a total of 10 features to predict stimulation responses.

\subsection*{Univariate analysis}

We began our analysis by investigating which of these features, individually, has the strongest association with stimulation responses. This analysis was carried out for each frequency band separately. We built univariate regression models relating each feature (separately) with each band-specific response across channels and sessions.

To account for the hierarchical nature of the data (observations from different channels within the same session and patient are not independent), we built linear mixed-effects models of the form $R \sim 1 + X + (1|S)$, where the independent variable $X$ represents any one of the ten features defined above. The term $(1|S)$ indicates that a random intercept was considered across groups of channels within a session, $S$. We consider random intercept models because it is important to detect common patterns of association across sessions, even if the average response varies from session to session.

\begin{figure}[t]
\captionsetup{width=1.4\textwidth}
\hspace{-45mm}\includegraphics{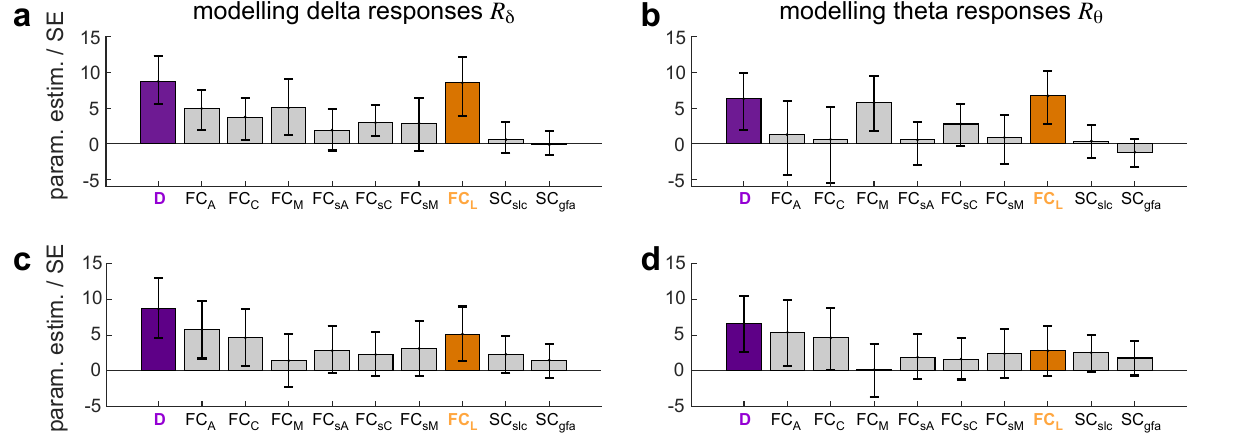}
\caption{
{\bf Univariate analysis of features.}
The associative power of each feature, given by the ratio between parameter estimate and its standard error (SE), was assessed by fitting the model $R \sim 1 + X + (1|S)$, where $X$ is one of the ten available features on the x-axis. The model considers a random intercept, with grouping based on sessions, as indicated by $(1|S)$. (a-b) The analysis was carried out on the subset of highly-responding sessions, which includes 20 sessions for delta and 25 sessions theta. The error bars correspond to the 95\% confidence interval produced through bootstrapping. The two strongest features for both delta and theta, $D$ and $FC_L$, are color-coded. (c-d) The same univariate analysis was carried out on the whole set of 131 sessions. Both $D$ and $FC_L$ continue to have high associative power.
}
\label{fig2}
\end{figure}

Any attempt of such session-grouped prediction relies on the assumption that there are some substantial stimulation responses in each session, at least in a handful of channels. We consider the responses to be substantial when they exceed the expected baseline fluctuations of band power. Indeed, our previous work showed that the majority of sessions in this dataset lack strong responses \cite{Papasavvas2020}. Thus, we focused on band-specific subsets of sessions with strong responses in at least two channels (see Methods). Furthermore, we consider only strong positive responses (i.e. increased band power) and not negative responses (i.e. decreased band power). This is due to the fact that strong positive responses are typically consistent across sessions (indicating a replicable biological effect) whereas strong negative responses are not \cite{Papasavvas2020}. 

The univariate analysis for each band was then applied to these subsets of highly responding sessions (20 sessions for delta, 25 sessions for theta). To quantify the association of each feature with the response we used a measure of associative power (AP): the feature's parameter estimate (coefficient) divided by its standard error, following previous work by Mohan and colleagues \cite{Mohan2020}. As expected, the inverse Euclidean distance $D$ was one of the strongest features in these subsets, while long-term variability of FC, $FC_L$, also showed a comparable associative power for both delta and theta (AP = 8.56 and AP = 6.72, respectively; Fig.~\ref{fig2}(a-b)). In fact, $FC_L$ was the feature with the highest association with theta responses, $R_\theta$. No substantial effect of $FC_L$ was found in the other frequency bands (see Ancillary files, Fig. S2). 

In order to validate that these associations were robust and did not rely on a few sessions, we ran a bootstrap process to produce the 95\% confidence interval of the associative power (error bars in Fig.~\ref{fig2}). The confidence intervals for both $D$ and $FC_L$ are above zero indicating the robustness of the associative power.

To generalize these results on delta and theta, we applied the same analysis across all available stimulation sessions (n=131) across 66 subjects, that is, including also sessions that did not show strong stimulation responses (Fig.~\ref{fig2}(c-d)). In both cases the inverse Euclidean distance $D$ was the feature with the highest associative power (AP = 8.69 for delta, AP = 6.57 for theta). However, $FC_L$ also exhibited a strong association (AP = 5.15 for delta, AP = 2.80 for theta) corroborating the results in Fig.~\ref{fig2}(a-b). For reference, an AP value above 2.5 approximately corresponds to a p-value below 0.01.

\subsection*{FC information collected over multiple days is required for the associative power of long-term variability of FC}

\begin{figure}[t]
\captionsetup{width=1.4\textwidth}
\hspace{-40mm}\includegraphics{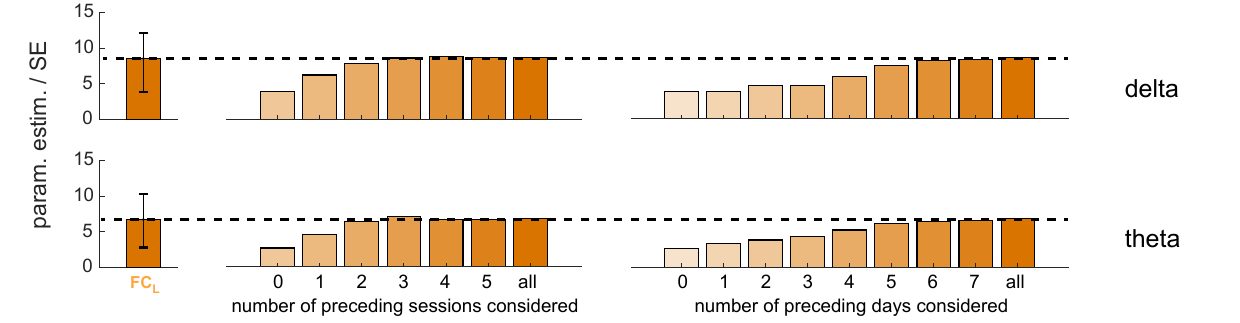}
\caption{
{\bf The associative power of long-term variability of FC increases with the gradual addition of FC information. }
The associative power of $FC_L$ on delta and theta responses increases almost monotonically as the calculation of FC variability includes an increasing number of preceding sessions. The middle panel shows this increase for every additional preceding session considered, regardless of its timing. Alternatively, the panel on the right shows the increase by lengthening the time-window from which preceding sessions were collected and included in the variability calculation. The time-window is gradually extended by a 24-hour step.  
}
\label{fig3}
\end{figure}

Next we looked closer at the strong association between $FC_L$ and the responses in delta and theta band (Fig.~\ref{fig2}). Since $FC_L$ is the only feature that summarizes FC information over multiple sessions and days, we asked how much information from preceding sessions and days is needed to achieve the strong associative power? 

We recalculated $FC_L$ and recomputed the univariate associations with an increasing number of preceding sessions included in the $FC_L$ calculation. The results shown in Fig.~\ref{fig3}a indicate that the AP increases rapidly with the first two additional preceding sessions while it plateaus with any further additions. This is the case for the univariate models of both delta and theta responses. Note that for zero preceding sessions, the $FC_L$ was calculated based only on the standard deviation between the anode and the cathode of the stimulation session under investigation (see again schematic in Fig.~\ref{fig1}d). 

An alternative analysis shows that the association increases steadily by gradually considering a longer time window from which we collected FC information. By including only the sessions in the preceding $K$ number of days, the association grew stronger with an increasing $K$ in both delta and theta. Figure~\ref{fig3}b shows that in order to maximize AP, we had to consider the sessions in the previous five to six days prior to the stimulation session under investigation.

This monotonic increase of associative power with the gradual extension of FC history was only found in delta and theta (for other bands, see Ancillary files, Fig. S3).

\subsection*{Step-wise analysis using linear mixed-effects models}

\begin{figure}[p]
\captionsetup{width=1.4\textwidth}
\hspace{-45mm}\includegraphics{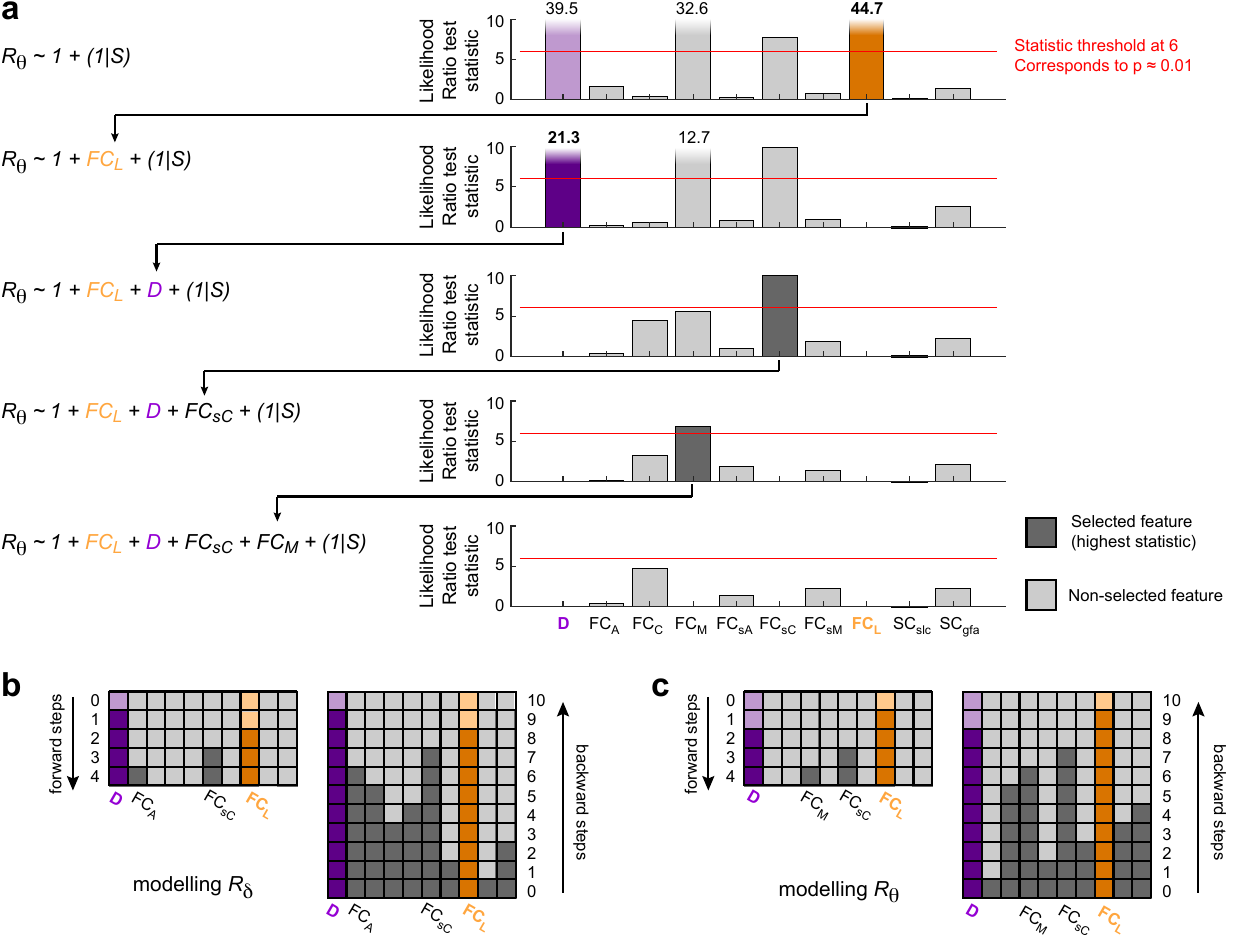}
\caption{
{\bf Step-wise analysis using multivariate mixed-effects models. }
(a) Model construction for theta. Starting from the constant model, the model was enriched by adding one feature at a time with Likelihood Ratio test statistic as the criterion. At each step the feature with the highest test statistic was added to the model. The process stopped when no feature remained with a statistic above 6 (red line; corresponds to $p \approx 0.01$). Here, any bar that surpasses 10 is truncated and explicitly labeled with its statistic. (b-c) Summary of both the forward and backward step-wise analysis for delta and theta, respectively. At each step (row), a feature was added to the model in forward analysis, or removed from it in backward analysis. Darker colors indicate features included in the model at each step. Forward step-wise analysis stopped when no more features could be added (due to statistic threshold), whereas backward analysis finished with the removal of the last feature.}
\label{fig4}
\end{figure}

To identify the most informative and complementary features, we followed a step-wise analysis process to build a linear mixed-effects model for the responses in each of the frequency bands. This analysis was applied on the subsets of highly-responding sessions, as before. Both forward and backward step-wise approaches were used and a Likelihood Ratio test was applied in every step to add or remove a feature from the model. The forward step-wise process that produced the model for theta responses is shown in Fig.~\ref{fig4}a. Notice that the process starts with a model without any features $R_\theta \sim 1 + (1|S)$ (i.e. constant model). Enriching features were added one by one, according to their ranking in improving model performance. The forward process was terminated when no additional feature could be found to substantially enrich the model. This enrichment was considered substantial as long as the Likelihood Ratio test statistic remained above 6, which approximately corresponds to a p-value of 0.01. The resulting model for theta responses is $R_\theta \sim 1 + FC_L + D + FC_{sC} + FC_M + (1|S)$. The features are ordered according to their importance.

Following the same process, the model produced for delta is $R_\delta \sim 1 + D + FC_L + FC_{sC} + FC_A + (1|S)$. Both forward and backward processes are summarized for delta and theta in Fig.~\ref{fig4}b-c, respectively. Notice that in both cases four features were added to the model with $D$ and $FC_L$ being the two most enriching features. It is evident that, in both delta and theta, the information provided by $D$ and $FC_L$ is complementary, since their respective statistic remained high even after the addition of the other feature. Finally, notice that the backward step-wise process verified the model produced by the forward process since it converged to the same four most important features (Fig.~\ref{fig4}b-c).

\subsection*{Prediction of responses using linear mixed-effects models}

After observing the strong associative power of $FC_L$ compared to other FC and SC features, and its complementary associative power to Euclidean distance, we formulated predictive models to evaluate whether $FC_L$ indeed boosts predictive performance in cross-validation tests.

In the following subsections we focus on formulating predictive models using linear mixed-effects models to predict the stimulation responses. For simplicity, we assessed the predictive power of models, which initially included a single feature $D$, and then we investigated the improvement of the prediction by adding a second feature. In other words we are testing predictive performance against the baseline model of only using Euclidean distance.

We performed cross-validation prediction at three levels. Through these levels, we gradually excluded an increasing amount of information from the training of the models on stimulation responses of the session on which prediction was performed. In other words, an increasing number of channel-specific responses in a session were held out from the training for a subsequent prediction. First, we performed prediction at the level of single channels; that is, predicting the response of a channel during a session given the responses of the remaining recorded channels, from the same session as well as the rest. Second, we performed prediction on half of the channels in a session, which could represent different combinations of grids and strips in the iEEG. Third, we performed prediction across all channels in a session, thus representing prediction on a new session for which we have no prior stimulation-related data. 

\subsection*{Prediction of the response at a specific channel}

\begin{figure}[p]
\captionsetup{width=1.4\textwidth}
\hspace{-45mm}\includegraphics{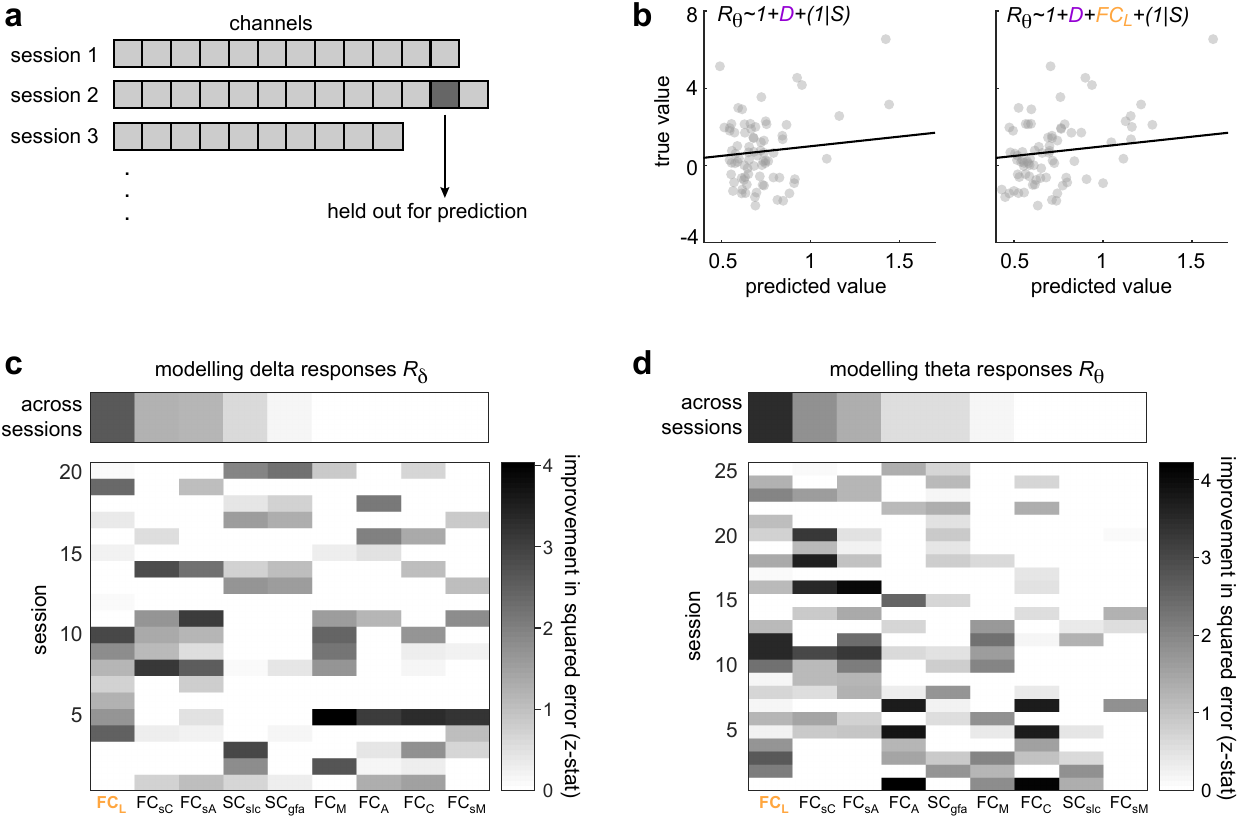}
\caption{
{\bf Prediction of stimulation responses at specific channels. }
(a) The response of each channel is predicted after training the model with the rest of the channels across all sessions. (b) Example of channel-wise prediction for a specific session. The true values are scattered against the predicted values produced by two different models of theta responses: the first featuring only the inverse Euclidean distance, $D$; the second having the additional feature of long-term variability of FC, $FC_L$. Line of equality is also shown. (c-d) Heat maps showing the prediction improvement based on the squared error between true and predicted responses, for delta and theta, respectively. The improvement (z-statistic of a paired non-parametric test) was assessed for different feature additions (x-axis) to the model $R \sim 1 + D + (1|S)$. The statistical test was applied to all the channels across sessions (top panel) but also to each session's channels separately (bottom panel). The features on the x-axis are sorted with descending improvement based on the results across sessions (top panel). Negative improvements (impairments) are set to 0 and shown in white for visual clarity.
}
\label{fig5}
\end{figure}

At this point, we want to explore whether the associative power of individual FC and SC features can be translated to meaningful response predictions for each individual channel. We investigate whether an additional feature can improve prediction over and above what is achieved by the model $R \sim 1 + D + (1|S)$. We used inverse Euclidean distance, $D$, as the first feature in the model due to its high associative power for delta and theta (Figs.~\ref{fig2}, \ref{fig4}), but also considering what is known from previous studies \cite{Solomon2018, Mohan2020}.

We systematically held out a channel while the rest of the data (remaining channels across all the highly responding sessions) were used for the models' training (Fig.~\ref{fig5}a). Each channel's response was predicted using two models: $R \sim 1 + D + (1|S)$ and $R \sim 1 + D + X + (1|S)$, where $X$ is an additional FC or SC feature. Illustrative predictions of theta responses with $FC_L$ as the additional feature, taken from an example session, are scattered against the true values in Fig.~\ref{fig5}b. An improved predictive performance can be seen, as the scattered points converge towards the line of equality (quantitatively, it yields a decrease in mean squared error from 2.72 to 2.54).

The squared error of each prediction (squared difference between true and predicted value) was calculated for both models. Then we quantified the prediction improvement as the z-statistic of a paired non-parametric test applied to these paired errors (see Methods). The strongest improvement was found with the addition of $FC_L$ to the model, for both delta and theta, as shown in Fig.~\ref{fig5}c-d , respectively. The upper panels show the improvement while considering all the channels across all sessions in each subset of highly-responding sessions (20 for delta and 25 for theta). The session-specific analysis in the lower panels reveals that the improvement is not due to a single session; but rather it is evident in the majority of sessions.

\subsection*{Prediction of responses in half of the channels in each session}

\begin{figure}[p]
\captionsetup{width=1.4\textwidth}
\hspace{-3mm}\includegraphics{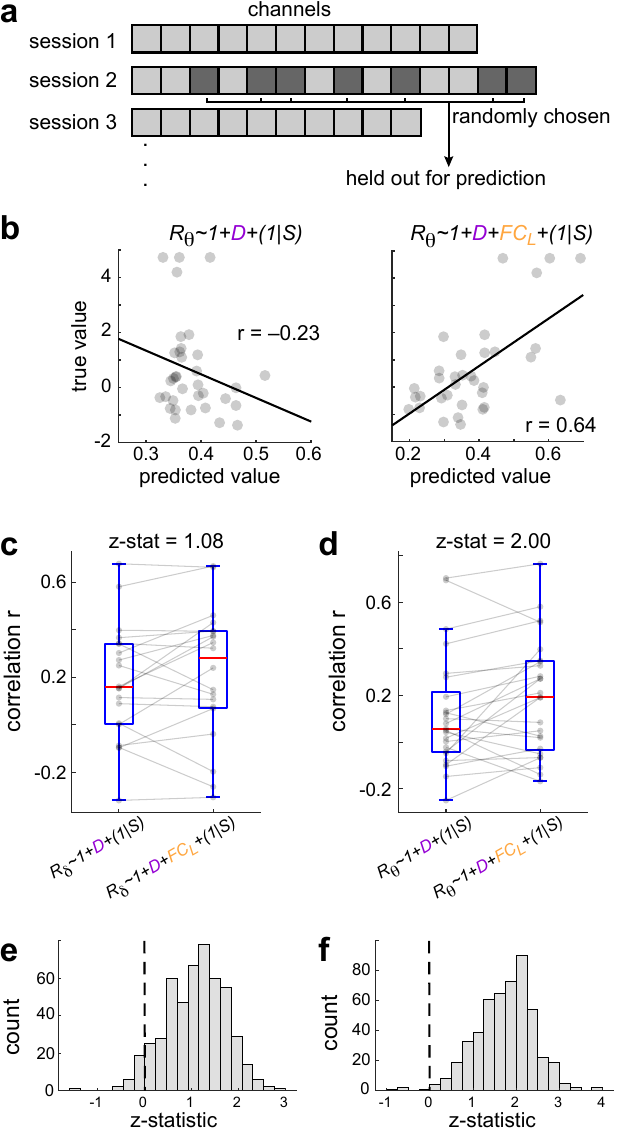}
\caption{
{\bf Prediction of stimulation responses in half of the channels in each session. }
(a) The predictive models $R \sim 1 + D + (1|S)$ and $R \sim 1 + D + FC_L + (1|S)$ were used to predict half of the channels in each session. For each session, the models were trained by all available channels except a randomly-chosen subset of half of the channels in the session, which were held out for prediction. (b) Illustrative examples of theta prediction, with the true responses scattered against the predicted ones. Least-squares line and correlation r between true and predicted responses are shown for each model. (c-d) Typical examples of prediction improvement across the highly responding sessions after including $FC_L$ in the models, for delta and theta, respectively. The z-statistic quantifying the improvement is shown at the top. (e-f) Distribution of z-statistics expressing the prediction improvement for delta and theta for 500 randomly-chosen subsets of channels.
}
\label{fig6}
\end{figure}

Since $FC_L$ exhibited the strongest predictive power on single-channel predictions, we analyzed it further by considering other cross-validation scenarios. It would be practically useful to predict subsets of channels in a session, for example, representing grids or strips in iEEG. In this subsection, we focus on the multi-channel prediction improvement after adding $FC_L$ to the model $R \sim 1 + D + (1|S)$. We wanted to test whether we can improve the prediction across multiple channels within a session by relying on information from the same session and the remaining sessions. We randomly held out half of the channels in each session and we predicted their responses after training the two models, with and without $FC_L$, on the rest of the channels across sessions (Fig.~\ref{fig6}a). This was repeated 500 times, thus applying prediction on 500 randomly selected sets of channels in each session. We first assessed the quality of the prediction by measuring the Pearson correlation r between the predicted responses and the actual responses in those channels (for an illustrative example on predicting theta responses, see Fig.~\ref{fig6}b). Session-specific correlation values were paired between the two models for each one of the 500 iterations. Figures~\ref{fig6}c-d show typical examples of the prediction improvement across all highly-responding session, for delta and theta respectively. Notice the typical increase of correlation with the inclusion of $FC_L$ to the model. The improvement was quantified with the z-statistic of a paired non-parametric test as before. The histograms of the  improvement values of the 500 repetitions are shown in Fig.~\ref{fig6}e-f. The prediction improvement in both delta and theta band is substantial with the inclusion of $FC_L$ (median z-stat=1.12 and median p=0.263 for delta; median z-stat=1.74 and median p=0.083 for theta).

Notice that the range of predicted values is typically much narrower than the range of true response values, as shown in the example of Fig.~\ref{fig6}b (and similarly in Fig.~\ref{fig5}b). This discrepancy is not important in our predictions since we aim in finding the channels more likely to respond and their relative differences, rather than their absolute responses. Thus, for the assessment of correlation, we used Pearson correlation which is insensitive to this discrepancy in ranges, since it normalizes out the range differences and relies on the relative rather than the absolute distances between the values (see Methods for more details).

\subsection*{Prediction of responses in across a whole session}

\begin{figure}[p]
\captionsetup{width=1.4\textwidth}
\includegraphics{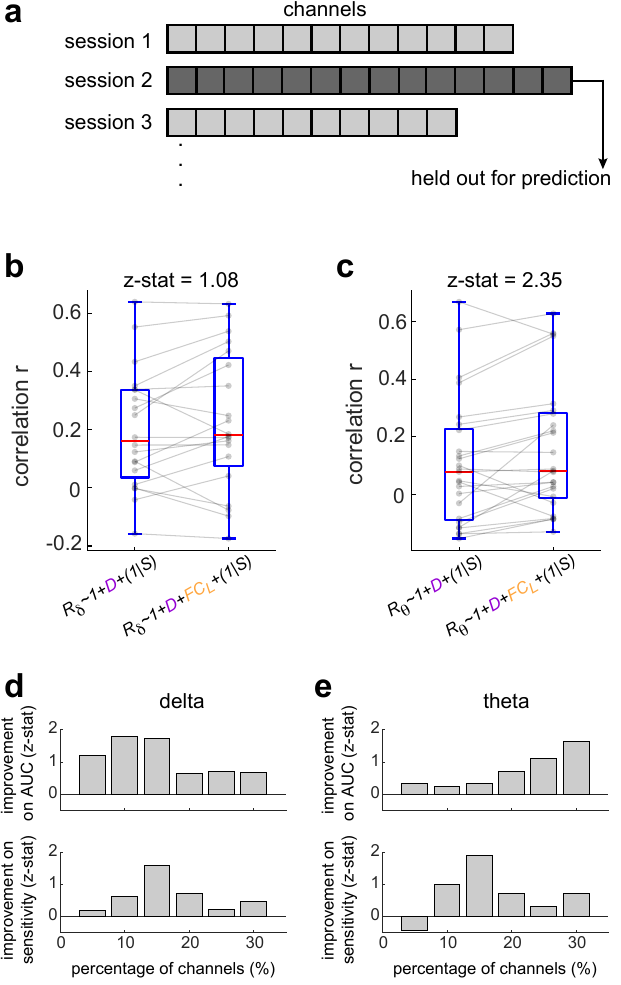}
\caption{
{\bf Prediction of stimulation responses across all channels in each session. }
(a) Cross-validation scheme for the prediction of whole sets of channels in a session. Each session was predicted by holding it out and training the models on the remaining sessions. (b-c) The correlations between predicted and true responses were calculated and the box plots summarize the distributions of those correlations, for delta and theta, respectively. The correlations tend to increase after the inclusion of $FC_L$ in the predictive model. The z-statistic expressing this improvement is shown at the top. (d-e) Improvement on two measures (Area Under the Curve and sensitivity) of binary prediction, where each channel was predicted to be either highly responding or not. The x-axis indicates the percentage of channels that were considered in each case to be highly-responding. The results are shown for delta and theta, respectively.
}
\label{fig7}
\end{figure}

At this final stage, we used the model to predict the responses for the different frequency bands across all channels in each one of the highly responding sessions, while training the model with the rest of the sessions (see Fig.~\ref{fig7}a for a schematic). The aim was to verify whether $FC_L$ can boost the prediction of whole sessions, following the encouraging results on half-sessions reported above. Once again, we investigated the improvement in prediction after introducing $FC_L$ to the model featuring only the inverse Euclidean distance. We computed the Pearson correlation between true and predicted responses for the two models: $R \sim 1 + D + (1|S)$ and $R \sim 1 + D + FC_L + (1|S)$. The correlation-based prediction improvement is shown in Fig.~\ref{fig7}b-c, for delta and theta, respectively. The improvement was modest for delta (z-statistic=1.08, p=0.279), while the prediction of theta responses was strongly improved (z-statistic=2.35, p=0.019).  

We then investigated whether we could predict the channels with the highest response in each session. That would be particularly useful in the context of neuromodulation protocols targeting specific brain regions. We classified each session's channels as expected to belong in the top Y\% or not, based on their predicted responses in a particular band. The predicted classes were then compared to the true classes, while we varied the percentage of channels considered to be the most responding channels from 5\% to 30\% (see Methods). We quantified the prediction improvement on two measures, namely the area under the receiver operating characteristic curve (AUC) and sensitivity, in the same way as we did for correlation (paired differences, see Methods). The results in Fig.~\ref{fig7}d show that the prediction of the top channels was generally improved for delta with the addition of $FC_L$, as evident by both AUC and sensitivity increases. Predicting the top 15\% of the channels for delta responses in each session was particularly improved through both measures. Similar results are shown in Fig.~\ref{fig7}e for theta, where improvements were found while predicting the top 10\% - 30\% of the channels in each session. Indicatively, the AUC was improved from 0.524 $\pm$ 0.138 to 0.615 $\pm$ 0.105 (median $\pm$ st. dev.) for the top 30\%, whereas sensitivity was improved from 0.125 $\pm$ 0.171 to 0.143 $\pm$ 0.206 for the top 15\%.

Finally, we investigated whether these findings are present even after focusing our analysis on each subject's first stimulation session. The purpose was to address the concern that the stimulation during a session can influence the long-term variability of FC in successive stimulation sessions. The results, found in the Ancillary files, suggest that the associative and predictive power of $FC_L$ remains high for both delta and theta even after excluding successive stimulation sessions from each subject.

\section*{Discussion}

Predictable stimulation responses are crucial for maximizing the therapeutic potential of brain stimulation. We showed that the stimulation-induced increase of delta and theta band power was predicted by long-term changes in FC, representing changes over multiple sessions and days. The predictive power of the long-term variability of FC relied on the strong association it exhibited with the corresponding responses; an association that matched the strong association of Euclidean distance and surpassed the associations of other FC and SC features tested (Fig.~\ref{fig2}). We demonstrated that the association gradually strengthened by increasing the amount of past history (in time and sessions) included in the calculation of long-term variability of FC (Fig.~\ref{fig3}). Finally, we showed that the predictive linear models for delta and theta responses were enriched and the predictions were substantially improved by adding long-term variability of FC to models already featuring the Euclidean distance. This suggests that the long-term variability of FC provides predictive information which is complementary to the spatial information provided by Euclidean distance.

We tested the predictability of responses through multiple cross-validation schemes: from single channels, to multiple channels, or even the whole set of channels in a session. This gradual testing of predictability enhances the validity of the results, demonstrating that long-term variability of FC contributes with useful information at every level. Furthermore, by looking at all these scenarios, we demonstrate the general applicability of such prediction in clinical practice. In some cases the goal is to predict the responses across all channels during an upcoming stimulation session, whereas in other cases we need to predict the responses at electrodes added later on. These diverse cross-validation strategies could guide the design of stimulation protocols in various situations.

The results suggest that there is a difference between the FC measures of anode and cathode, $FC_A$ vs $FC_C$ and $FC_{sA}$ vs $FC_{sC}$, and their relationship with the responses. Despite the fact that none of these measures showed a particularly strong associative or predictive power compared to $FC_L$, they still carry information which is not necessarily shared between anode and cathode (see Figs.~\ref{fig2}, \ref{fig4}, and ~\ref{fig5}). For instance, the short-term variability of cathode, $FC_{sC}$, exhibited much stronger association with theta responses than the short-term variability of anode, $FC_{sA}$ (see Fig.~\ref{fig4}a). Despite the bipolar nature of the stimulation, anode and cathode were specifically annotated in the metadata, possibly indicating which electrode had the leading phase during each biphasic pulse. Recent investigations on the relationship between stimulation effect and pulses' leading phase showed that cathode-leading stimulation activates more neurons across a wider area compared to anode-leading stimulation \cite{Stieger2020, Park2018}. Thus, the observed differences in our results could be due to differential responses of the tissue between anode and cathode site despite the bipolar nature of stimulation. 

The observed channel-specificity in long-term FC changes might indicate functional reorganization in specific edges of the network. Specific functional connections between stimulation and responding sites were found to undergo slow changes over days and the same responding sites were the ones responding strongly to stimulation. A recent experimental study might provide a potential explanation of this observation. Lesion-induced reorganization in rodents was associated with hyper-excitability of brain areas that had been involved in such reorganization \cite{Verley2018}. Network reorganization may underlie functional recovery or compensation \cite{Mbwana2009}. Intra- and inter-hemispheric functional reorganization has been reported before in patients \cite{Mbwana2009, Chang2017, Wirsich2016}. Some of the observed slow changes in our analysis may represent functional reorganization due to the epilepsy.

Another potential explanation of our main finding may involve maintenance mechanisms of plasticity. Huang and colleagues reported an association between highly responding channels during stimulation and changes in effective connectivity, which indicate the maintenance of induced plasticity minutes after stimulation \cite{Huang2019}. Our results reinforce the potential association between increased excitability and plastic changes in functional and effective connectivity.

Associations between structural connectivity and neuromodulatory responses are still elusive. Multiple studies have reported predictability of clinical outcomes based on the structural connectivity of the stimulation site within a pathology-related network \cite{Horn2017, Johnson2020, Riva-Posse2014, Coenen2019}. However, none of these studies provided evidence that the clinical improvement was mediated through the stimulation-induced modulation of band power in anatomically connected regions. Despite the fact that neuromodulation has been reported to be stronger when stimulation is delivered near white matter \cite{Mohan2020, Solomon2018}, this does not necessarily translate to association between neuromodulation at the response sites and their structural connectivity to the stimulation site. The present study might be the first one analyzing band power modulation at anatomically-connected distant sites by considering the structural connectivity of the stimulation site. The structural connectivity measures used here exhibited weak association with the stimulation responses in general. A possible caveat, and limitation of this study, is that the structural connectivity was estimated from healthy controls not the individual patients who experienced stimulation. Ideally, future work should use subject-specific structural connectivity data to investigate whether SC can provide predictions for neuromodulatory effects on activity's band power.

Furthermore, a future investigation could focus on the prediction of the presence of highly responding channels in a session. Our predictability analysis here involves only those stimulation sessions which include strong responses, in a band-specific way, and the prediction is applied on each channel. However, it is still unclear whether it is possible to predict those sessions in the first place. Arguably, features characterizing the brain state during stimulation would serve better in such predictions \cite{Jensen2011, Li2019, Ruhnau2016, Chiang2021}. Indeed, the brain state has already been found to be an enabling factor for effective neuromodulation through stimulation (for a review on state-dependent stimulation \cite{Jensen2011}). Ongoing oscillatory activity and cognitive tasks that activate distributed brain networks can determine the effectiveness of stimulation \cite{Ruhnau2016, Li2019}. Extracting characteristics of the brain state during the session's initial period could potentially provide predictive information on brain's responsiveness during stimulation. 

It would also be interesting to explore whether these findings can be verified in other datasets or recording modalities. Unfortunately, datasets with multiple sessions across multiple days are scarce, especially recordings of intracranial EEG. However, in principle, these findings could be observable in noninvasive EEG studies with transcranial direct current stimulation (tDCS). Since we are interested in responses in delta and theta bands, non-invasive scalp EEG could detect those responses, while the FC can be monitored through multiple EEG recording sessions over days.

In summary, we have shown that slow changes in functional connectivity provide predictive power over the stimulation responses in delta and theta band power. Through multiple cross-validation schemes, we demonstrated the potential applicability of this predictive power in clinical practice. Further research is needed to determine how this association emerges. Its understanding will maximize the impact of functional network dynamics in therapeutic applications of brain stimulation.

\nolinenumbers


\section*{Ancillary files}
This preprint is supported by Ancillary files found at arXiv:2105.02805.



\bibliographystyle{unsrt} 
\bibliography{myBib.bib} 


\end{document}